\documentclass[dvipsnames]{article}

\usepackage[margin=1in]{geometry}

\usepackage{graphicx,amsmath,upgreek,bm}
\usepackage{colortbl}
\usepackage{wrapfig}
\usepackage[rgb, table]{xcolor}
\usepackage{booktabs}
\usepackage{makecell}
\usepackage{pifont}
\usepackage{todonotes}
\usepackage[toc,page]{appendix}
\usepackage{multirow}
\usepackage{booktabs}
\usepackage{algorithm}
\usepackage{enumitem}
\usepackage{appendix}
\usepackage{color}
\usepackage{hyperref}
\hypersetup{
   colorlinks=true,
   linkcolor=blue,
   filecolor=magenta,
   urlcolor=blue,
   citecolor=blue,
	}
\usepackage[
  separate-uncertainty = true,
  multi-part-units = repeat
]{siunitx}
\usepackage{caption} 
\usepackage{subcaption}
\usepackage{amsmath}
\usepackage[noend]{algpseudocode}
\usepackage{stackengine}
\usepackage{multirow}
\usepackage{arydshln}
\makeatletter
  \renewcommand*\env@matrix[1][*\c@MaxMatrixCols c]{%
    \hskip -\arraycolsep
    \let\@ifnextchar\new@ifnextchar
  \array{#1}}
\makeatother
\makeatletter
\def\BState{\State\hskip-\ALG@thistlm}
\makeatother
\definecolor{light-gray}{gray}{0.9}

\title{On Transfer Learning of Traditional Frequency and Time Domain Features in Turning}

\author{Melih C.~Yesilli\\
				Department of Mechanical Engineering\\
				Michigan State University\\
				yesillim@egr.msu.edu
			\and
				Firas A.~Khasawneh\\
				Department of Mechanical Engineering\\
				Michigan State University\\
				khasawn3@egr.msu.edu
}
				
\date{}

\begin{document}

\maketitle    

%%%%%%%%%%%%%%%%%%%%%%%%%%%%%%%%%%%%%%%%%%%%%%%%%%%%%%%%%%%%%%%%%%%%%%
\begin{abstract}
There has been an increasing interest in leveraging machine learning tools for chatter prediction and diagnosis in discrete manufacturing processes. 
%The idea is to use tagged data obtained from sensor measurements to train a classifier so that an incoming data stream can then be tagged as chatter or chatter-free. 
%This process is referred to as supervised learning, and its success is strongly dependent on extracting features indicative of chatter from a training data set. 
Some of the most common features for studying chatter include traditional signal processing tools such as Fast Fourier Transform (FFT), Power Spectral Density (PSD), and the Auto-correlation Function (ACF). 
%These tools have been used successfully in several experiments to identify the occurrence of chatter, typically through visually comparing the time and frequency signatures against the machine tools' chatter frequencies. 
%However, even if a classifier gives high accuracy rates for a certain set of cutting parameters, its implementation on an actual cutting center is also contingent upon the robustness of the classification to inevitable shifts in the process parameters such as temperature changes or variation in the structure's natural frequencies. 
In this study, we use these tools in a supervised learning setting to identify chatter in accelerometer signals obtained from a turning experiment. 
The experiment is performed using four different tool overhang lengths with varying cutting speed and the depth of cut.
We then examine the resulting signals and tag them as either chatter or chatter-free. 
The tagged signals are then used to train a classifier.
%, and the resulting accuracies are tested by splitting the data into training and testing sets. 
The classification methods include the most common algorithms: Support Vector Machine (SVM), Logistic Regression (LR), Random Forest (RF), and Gradient Boost (GB). 
Our results show that features extracted from the Fourier spectrum are the most informative when training a classifier and testing on data from the same cutting configuration yielding accuracy as high as $\%96$. 
However, the accuracy drops significantly when training and testing on two different configurations with different structural eigenfrequencies. 
Thus, we conclude that while these traditional features can be highly tuned to a certain process, their \textit{transfer learning} ability is limited. 
%This makes these methods suitable for manufacturing processes whose sensory signals remain close to the ones used for training. 
%Alternatively. if a big shift is expected in the time series obtained from the sensors, then care must be taken when using any FFT/PSD/ACT based features. 
We also compare our results against two other methods with rising popularity in the literature: Wavelet Packet Transform (WPT) and Ensemble Empirical Mode Decomposition (EEMD). 
The latter two methods, especially EEMD, show better transfer learning capabilities for our dataset. 
\end{abstract}


\section{Introduction}
\label{sec:Intro}
% Machine tool chatter is one of the main causes of reduced productivity in discrete manufacturing operations. 
One of the essential problems in machining processes such as milling and turning is the occurrence of excessive cutting tool vibration called chatter. 
While there is over a decade long history of chatter literature, the practical implementation of many of the existing theoretical and numerical results remains a challenging task. 
This is caused by the significantly more complex nature of industrial processes in comparison to the simplified models in the literature. 
Therefore, in addition to predictive models, which are important for guiding the manufacturing process, machine learning has emerged as a tool for chatter diagnosis from sensors mounted on the cutting center. 
The essence of these tools is to extract features from the signal that capture the main characteristics of chatter, and train a classifier that can hopefully identify chatter in new data streams. 
% Chatter prediction and the mitigation are main focus of the studies in the literature. 
% There several chatter mitigation techniques such as increasing the stiffness of the cutting tool, increasing the damping and changing the spindle speed \cite{Munoa2016,Altintas1992, Yilmaz2002,2003alregib}. 

There are several ways for extracting features indicative of chatter from sensor signals. 
Two of the most widely used methods in the literature are based on the Wavelet Transform (WT) \cite{Yao2010,Cao2013,Yoon2010,Saravanamurugan2017,Qian2015,Chen2017, Zhang2016b,Choi2003,Yesilli2019} and the Empirical Mode Decomposition (EMD) \cite{Yesilli2019,Chen2018,Liu2011,Ji2018}. 
Both of these methods require the user to perform manual preprocessing to identify the wavelet packet or decomposed signals which includes chatter frequency.
Then, time and frequency domain features are pulled from these packets or decompositions and are used in classification algorithms. 
When the number of extracted features is large, the feature extraction tools are combined with feature ranking algorithms to determine a smaller subset of the most informative features in both frequency and time domains \cite{Chen2017,Yan2015,Yesilli2019}
Although Wavelet Transform and EMD are widely used methods, they have some shortcomings such as manual preprocessing of the data and requiring new sets of data for a new cutting configuration. 
For example, new cutting configurations may lead to a shift in the natural frequencies which in turn can change the chatter frequencies thus necessitating the repeating the classifier training in order to extract new informative features form the signal. 
However, it is seldom the case that the cutting parameters remain the same throughout the machining process. 
Therefore, one practical approach is to search for features that can capture chatter features that transcend deviations in the process, and therefore, can still yield reasonable accurate results when used to train a classifier on one set of parameters \cite{Yesilli2019}, and test it on a different set---a concept commonly referred to as \textit{transfer learning}. 
There are also several recent methods for chatter detection based on topological features in the signal \cite{Yesilli2019a,Khasawneh2016,Khasawneh2018, Yesilli2019c}, or on similarity measures via Dynamic Time Warping (DTW) combined with K-Nearest Neighbor algorithms \cite{Yesilli2019b}. 
Alternatively, deep learning can be used for chatter detection albeit with sufficiently large training sets \cite{Fu2015,Lamraoui2013,Kuljanic2009,Tansel1991}.

We remark that the number of applications on transfer learning has been increased during the last two decades. 
While some of these applications are related to machining processes \cite{Yesilli2019}, others include WiFi localization \cite{Pan2007, Pan2008,Zheng2008}, human activity recognition \cite{Harel2010}, identifying the burn severity of the forest fires \cite{Zheng2020}, software defect detection \cite{Nam2013,Nam2015}, natural language processing \cite{Blitzer2006,Houlsby2019,Ruder2019}, speech recognition \cite{Kunze2017}, music classification \cite{Wang2019}, and acoustic scene classification \cite{Mun2017}.
Self-taught learning results were also shown for image classification, character recognition, song genre, web page and article classification \cite{Raina2007}. 

All of the above methods for feature extraction in chatter are based on advanced concepts from files such as signal processing, topological data analysis, or statistics. 
However, surveying the machine tool chatter literature reveals that some of the main tools that researchers utilize to identify chatter, not necessarily within machine learning context, are traditional signal processing methods such as the Fast Fourier Transform (FFT), Power Spectral Density (PSD), and the Auto-correlation Function (ACF). 
Very few of these studies combine these features with machine learning. 
For example, Potocnik et al.~used the amplitude of the peaks in the power spectra to detect chatter in band sawing process \cite{Potocnik2013}. 
FFT has been also used with PCA as feature extraction method for classification of EMG signals \cite{Gueler2005}.  
Besides the limited number of works that combine these features with machine learning, these tools are often utilized for visual inspection, for example, of peaks near the natural frequencies of the tool or the structure. 
However, there is no standard approach for incorporating these traditional features into a systematic machine learning framework for chatter detection. 
Further, it is not clear how conducive to transfer learning these features are. 

In this paper, we use traditional signal processing methods to extract features from an accelerometer attached to the tool post of a lathe.  
The features used for training a chatter classifier are the coordinates of the peaks in the signal's FFT, ACF, and PSD. 
We overcome the overfitting problem resulting from the large number of features by combining the classification algorithms with Recursive Feature Elimination (RFE)---a feature ranking approach. 
We note that there are other methods for feature selection and reducing the number of features such as Principal Component Analysis (PCA) \cite{Song2010}.
We employ the four most widely used classification algorithms: Support Vector Machine (SVM), Logistic Regression (LR), Random Forest Classification (RF), and Gradient Boosting (GB). 
We compare the resulting accuracies for the example data set from FFT, ACF, and PSD features to their WPT and EEMD counterparts. 

The paper is organized as follows. 
Section \ref{sec:Transfer_Learning} provides technical details of transfer learning.
Section \ref{sec:Experimental Setup} explains the experimental procedure and data acquisition for the turning experiment. 
Section \ref{sec:Feature_Extraction} describes the feature extraction methods and the feature ranking method. 
Section \ref{sec:Results} provides the results, while Section \ref{sec:Discussion} includes the discussion. 
We provide the concluding marks in Section \ref{sec:Conclusion}.

\section{Transfer Learning}
\label{sec:Transfer_Learning}
\begin{figure}[h]
\centering
\includegraphics[width=0.5\textwidth]{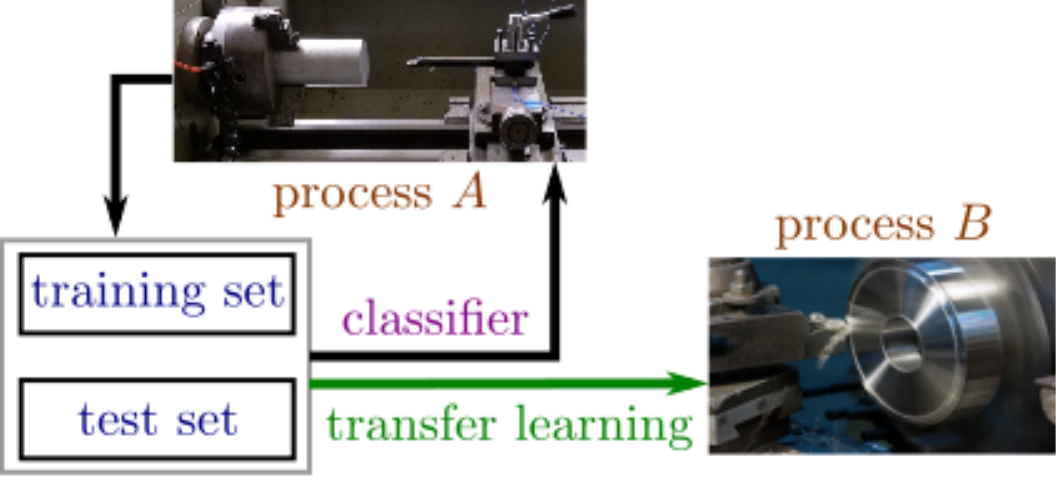}
\caption{Transfer learning is concerned with extending knowledge learned from one system to another system.}
\label{fig:transfer_learning_schematic}
\end{figure}
For most of the tasks, the human learning process is cumulative in nature. 
We often transfer our previous knowledge and experiences to learn new tasks and that make our life easier and more efficient. 
This concept can be extended to machine learning algorithms with the help of transfer learning. 
Specifically, traditional machine learning algorithms requires both test and training set data to be composed of the same feature space or to have the same distribution \cite{Pan2010,Dai2007}.
However, this is difficult to achieve in real life applications, such as chatter detection, because there many parameters that affect the experiment. 
These parameters can vary from run to run, or even within the same run due to the tool center movement, heat effects, and ambient vibrations. 
If the traditional machine learning approach is to be applied, this means that data will need to be collected for each parameter variation; thus leading to time consuming and expensive effort. 
Therefore, it is advantageous to develop machine learning algorithms that need to be trained on a certain training data but can be tested on systems with cutting configurations that deviate from the system that was used for training. 
\textit{Transfer learning} makes this transfer of knowledge possible thus allowing more flexibility in the source of the trained classifier and its target, see Fig.~\ref{fig:transfer_learning_schematic}. 

\begin{figure*}
\centering
\includegraphics[width=0.95\textwidth]{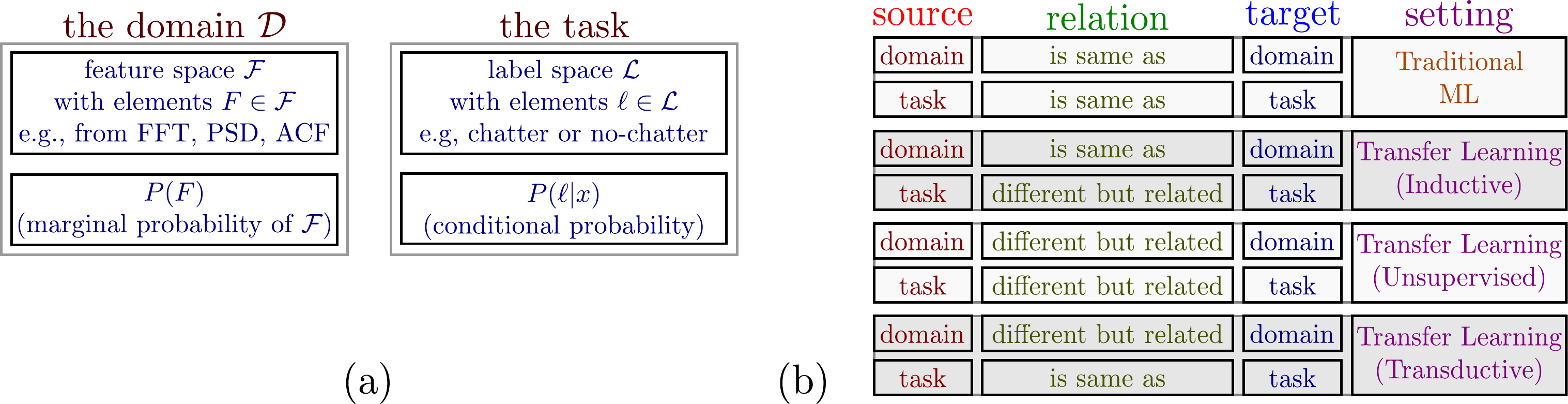}
\caption{(a) Elements of the domain and the task. (b) Classification of different machine learning settings. }
\label{fig:transfer_learning_cat}
\end{figure*}

The definition and categorization of transfer learning is facilitated by introducing the concepts of tasks and domains \cite{Pan2010}, see Fig.~\ref{fig:transfer_learning_cat}a. 
The domain contains the feature space $\mathcal{F}$ with features $F \in \mathcal{F}$, as well as the marginal probability of the features $F$. 
The task is composed of the label space, and the conditional probability. 
Using these definitions, we can think of two systems each with their own domain and task: a source which is the system that we are using to train the classifier, and a target which is the system we want to which we are applying the classifier. 
Depending on the relationship between the domains and the tasks of the source and the target, we obtain different machine learning settings \cite{Pan2010}, see Fig.~\ref{fig:transfer_learning_cat}. 
Out of the different categories in Fig.~\ref{fig:transfer_learning_cat}, inductive transfer learning is of particular interest to this paper. 
In this setting, the domain is the same for the source and the target, and the tasks are related but different. 
This setting of transfer learning is further separated into two different categories: multi-task learning and self-taught learning. 
In multi-task learning, both source and target domains have labeled data. 
In contrast, in self-taught learning labeled data is not present in the source domain. 
When it is said that domains or the tasks are different between source and target, there is a relation between them. 
Otherwise, there will not be any information to transfer and it can lead to $negative$ $transfer$ which shows detrimental effect on the classification results for the target task \cite{Weiss2016}.
For more detailed information about each category, see Ref.~\cite{Pan2010, Weiss2016, Lu2015}.
In this paper, our source and target are represented as turning processes with cutting configurations that have different overhang lengths.  
Therefore, the application of transfer learning in our work fits into inductive transfer learning category since we have different, but related, tasks for the source and target systems. 

% In addition, transductive transfer learning has the opposite settings of inductive transfer learning. 
% The source and targets are the same in this case. 
% Transductive transfer learning also requires to have unlabeled data set in the target domain.
% The last category is the unsupervised transfer learning where the labeled data is not available for both source and target domain. 
% The domains and tasks for source and target are different for unsupervised transfer learning. 

% Wang and Mahadevan proposed manifold alignment approach to transfer knowledge between different feature spaces \cite{Wang2011}.

% The label space ($\mathcal{L}$) and the conditional probability distribution ($P(l|x)$) is the elements of the task for a specific domain. 
% The label space ($\mathcal{L}$) represent the all labels, while $l$ is the element of the label space. 
% In our application, we can define the label space as the space including labels of existence of chatter or chatter free. 
% Our feature space is composed of features extracted from FFT, PSD, and ACF. 

\section{Experimental Procedure}
\label{sec:Experimental Setup}
The cutting tests were performed by varying the overhang length of the cutting tool, the rotational speed of the spindle, and the depth of cut. 
The equipment used in the experimental setup are listed in Table~\ref{tab:Equipment_list}.

\begin{minipage}{\textwidth}
  \begin{minipage}[a]{0.45\textwidth}
    \centering
    \includegraphics[width=1\textwidth,height=.95\textheight,keepaspectratio]{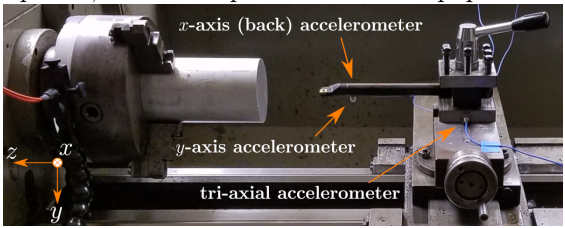}
    \captionof{figure}{Experimental Setup}
	\label{fig:Experimental_Setup}
  \end{minipage}
  \hfill
  \begin{minipage}[b]{0.53\textwidth}
    \centering
	\captionof{table}{Equipment list}
	\label{tab:Equipment_list}
	\resizebox{0.95\textwidth}{!}{	
	\begin{tabular}{cl}
	\hline
	Item Number & Equipment\\
	\hline
	1						& Clasuing-Gamet 33 cm (13 inch) engine lathe\\
	2 					& PCB 352B10 uniaxial accelerometers\\
	3						& PCB 356B11 triaxial accelerometer \\
	4						& NI USB-6366 data acquisition\\
	5 					& Titanium nitride coated insert\\
	6 					& S10R-SCLCR3S boring bar\\
	\hline
	\end{tabular}}
  \end{minipage}
  \end{minipage}

Two uniaxial accelerometers were attached to the boring bar, and one tri-axial accelerometer was placed on the tool holder, see Fig.~\ref{fig:Experimental_Setup}. 
Only the $x$ axis acceleration signals from the tri-axial accelerometer was used in the subsequent analysis since we found it had the best signal-to-noise ratio. 
No in-line analog filter was used; instead, the data was oversampled at 160 kHz and a Butterworth low pass filter with order of 100 was used before downsampling the data to 10 kHz, which is the top of the useful range of the used accelerometer.

\begin{figure}[h]
\centering
\includegraphics[width=0.3\textwidth,height=.5\textheight,keepaspectratio]{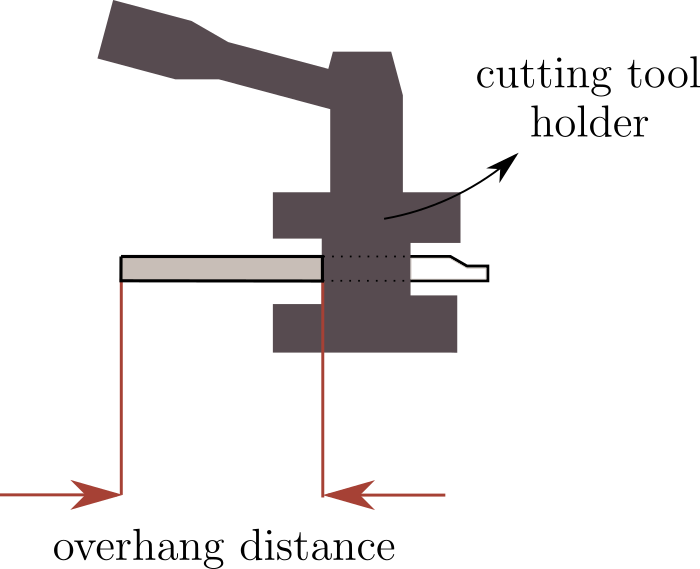}
\caption{Illustration of the overhang length.}
\label{fig:overhang_length}
\end{figure}

The cutting tests were performed for four different overhang lengths, which is the distance between the back of the boring bar and the heel of the cutting tool holder, see Fig.~\ref{fig:overhang_length}. 
These lengths are 5.08 cm, 6.35 cm, 8.89 cm and 11.43 cm (2 in, 2.5 in, 3 in, and 3.5 in, respectively)
For each overhang length, several experiments were performed by varying spindle speed and the depth of cut.
Then, each time series was analyzed and labeled as no-chatter (stable), intermediate (mild) chatter, or chatter (unstable) for supervised classification.
This tagging was performed by analyzing the signals in both time and frequency domains.  
Signals with low amplitudes in time and frequency was labeled stable (chatter-free). 
Data with low amplitude in time, but large amplitude in frequency was tagged intermediate (mild) chatter. 
But if the signal had large amplitudes in both domains, they were tagged as chatter. 
Signals that did not fit any of the above descriptions were tagged as unknown, and they were excluded from the analysis. 
Although we tagged intermediate and full chatter as two separate classes, in this study we combined chatter and intermediate chatter into one class, and performed two-class classification: chatter and no-chatter. 
We posted the raw data, including signals from all the mounted sensors, as well as the full tagging in the Mendeley repository \cite{Khasawneh2019}.
%------------------------------
%------------------------------

\section{Feature Extraction}
\label{sec:Feature_Extraction}
\begin{figure}[t]
\centering
\includegraphics[width=0.6\textwidth,height=.95\textheight,keepaspectratio]{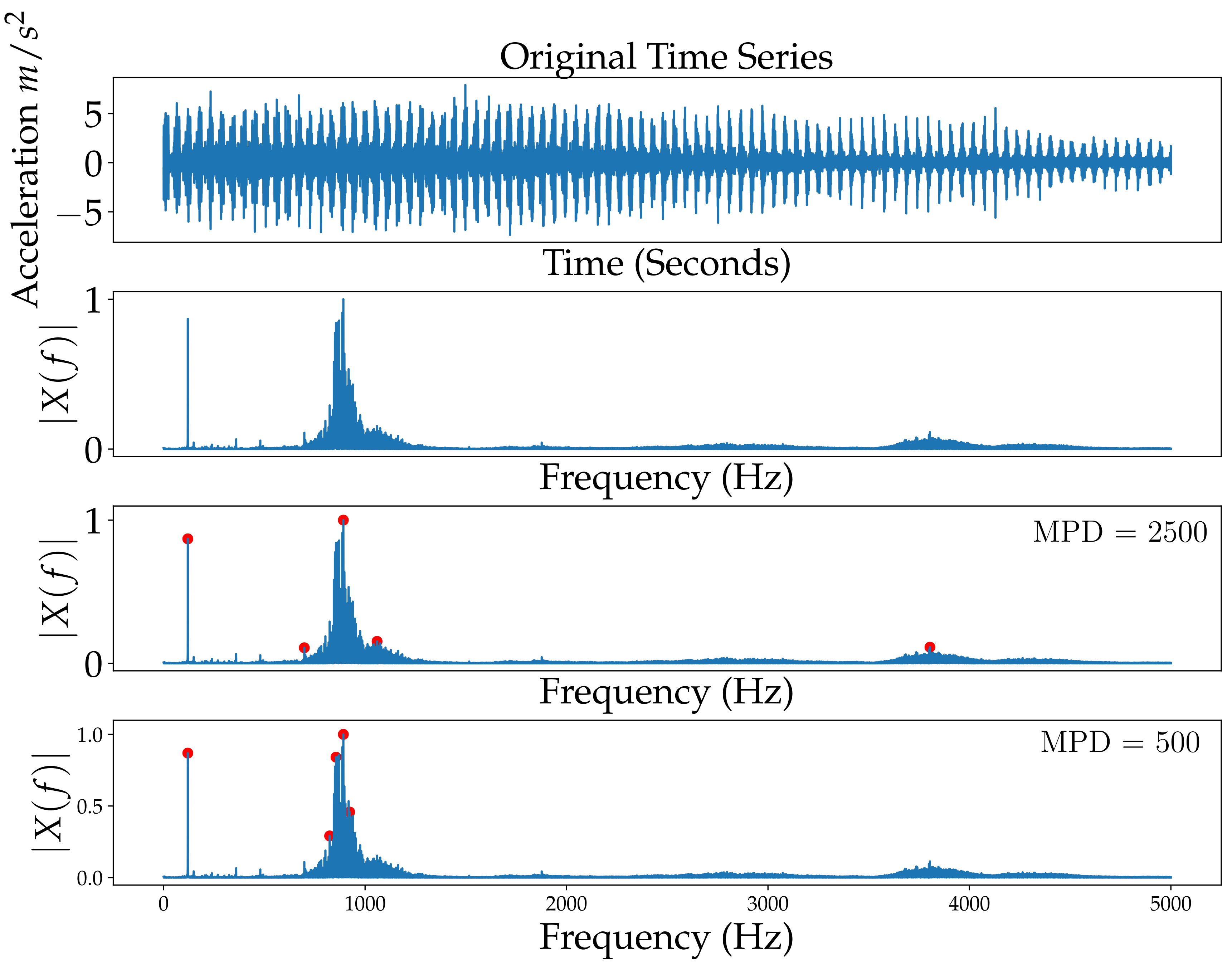}
\caption{First five peaks in Fast Fourier Transform (FFT) plot of the time series of acceleration signal collected from cutting configuration with 5.08 (2 in) cm overhang length, 320 rpm rotational speed of spindle and 0.0127 cm (0.005 in) depth of cut. (Minimum Peak Distance: 500 and 2500)}
\label{fig:FFT_example}
\end{figure}

In this section, we briefly describe how we extracted a feature vector composed of elements from the three traditional signal processing methods (FFT, PSD, and ACF). 
Features for FFT, PSD and ACF are obtained by using their peaks' coordinates.  
We used as features both the $x$ and $y$ components of the first five peaks in each of these three functions. 
Although there are some built-in commands for peak finding in the most common scientific software tools, reliable peak selection remains a challenging task. 
This is due to the large number of returned `bumps' which correspond to local maxima that are artifacts of noise, and are not true features of the signal.   
Therefore, we need to impose some constraints to sift out the redundant peaks and capture the most useful ones. 
The peaks of the FFT, PSD and ACF are selected by defining minimum peak height and the minimum distance between two consecutive peaks. 
The minimum peak distance was kept constant for all sequences while the minimum peak height is computed as a fraction of the difference between the $5$th and the $95$th percentile of the peaks according to
\begin{equation*}
MPH = y_{\rm min}+ \alpha(y_{\rm max}-y_{\rm min}), \text{ where } \alpha \in [0, 1],
\end{equation*}
and $y_{\rm min}$ is the $5$th percentile of the amplitude of the FFT/PSD/ACF, while $y_{\rm max}$ is the $95$th percentile. 
Fig.~\ref{fig:FFT_example} provides the original cutting signals along with its spectrum including the first five peaks. 
In Fig.~\ref{fig:FFT_example}, we provide two sets of first five peaks based on chosen minimum peak distance (MPD) parameter. 
The figure shows that $MPD=500$ chooses more points near the maximum amplitude in comparison to $MPD=2500$.
However, some of the cutting signals spectra do not contain five peaks if $MPD=2500$ is used. 
Therefore, we set $MPD=500$ to consistently extract five peaks from the FFT of all signals, but we used $MPD=1000$ with the auto-correlation function. 
Because the power spectral density plots were smooth, we did not use an MPD constraint with them.   

The feature matrices were given as input to four different supervised machine learning techniques: Support Vector Machine (SVM), Logistic Regression (LR), Random Forest Classification (RF) and Gradient Boosting (GB).
Vibration signals for each overhang size were split into $\%67$ training set and $\%33$ test set. 
Recursive Feature Elimination (RFE) was then used to rank the features. 
Splitting the data and performing classification was performed ten times. 
The resulting mean accuracies and standard deviations are provided in Sec.~\ref{sec:Results}. 
 
%----------------------------RFE---------------
\subsection{Recursive Feature Elimination (RFE)}
Classification algorithms are combined with Recursive Feature Elimination (RFE). 
RFE is iterative process that eliminates one feature in each iteration and provides a rank list for the features.  
The effect of each feature on classification determines which feature is eliminated in each iteration \cite{Guyon2002}. 
After ranking the features, we generated feature vectors such that the first feature vector includes only the first ranked feature and the next one adds the next feature in the list until all the features are used in the ranking list. 
For five different peaks, we have $30$ features: the $x$ and $y$ coordinates of the peaks in each of the FFT, ACF, and PSD. 
This resulted in $30$ different feature vectors according to the ranking list. 
Then, we decreased the total number of features for a signal by decreasing the number of peaks used in feature extraction from five to two.
Therefore, we ended up with $12$ features. 
These features are ranked and they are used for training the classifiers.

\section{Results}
\label{sec:Results}

\begin{figure}[H]
\centering
\includegraphics[width=0.9\textwidth,height=.95\textheight,keepaspectratio]{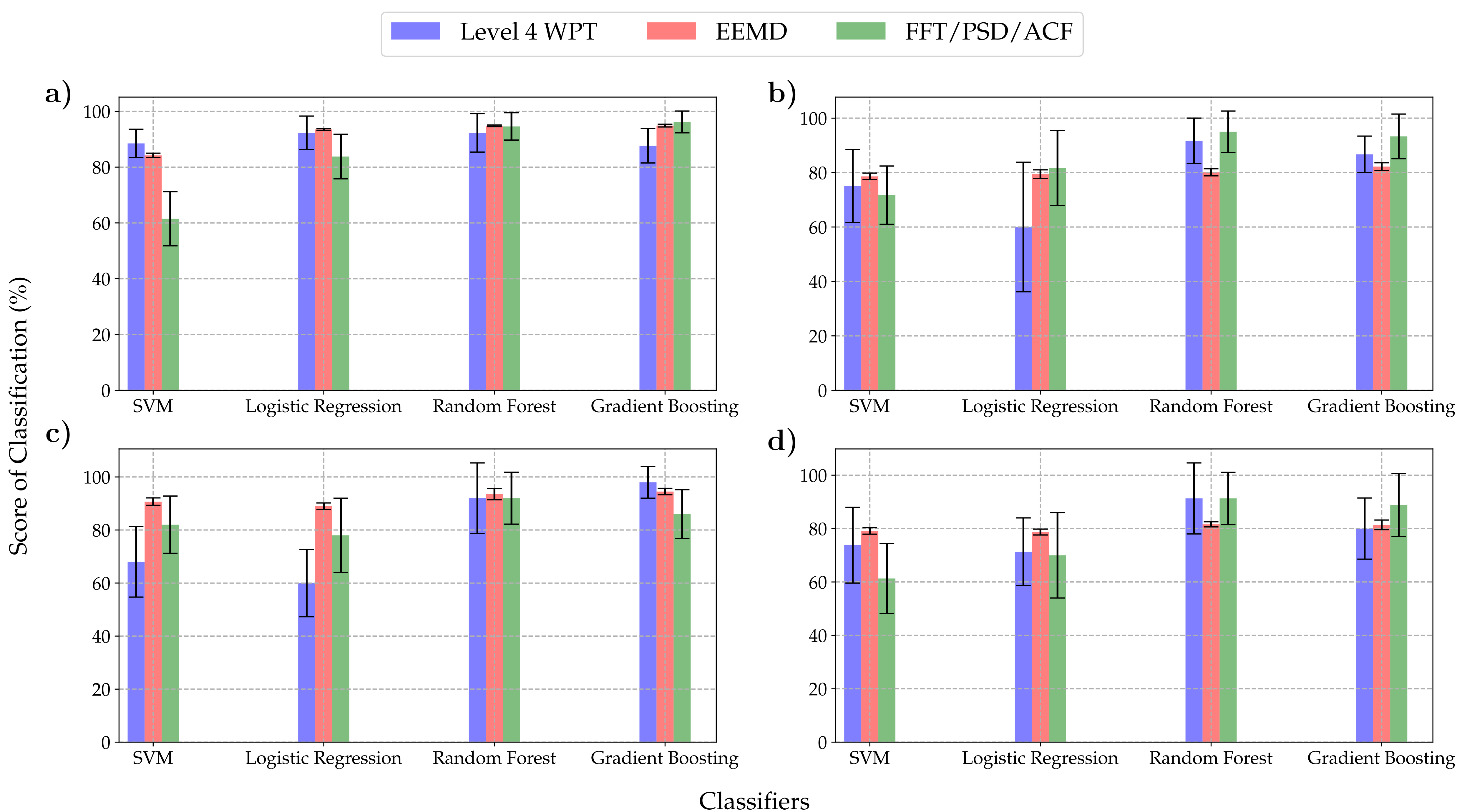}
\caption{Comparison of classification results of Level 4 WPT, EEMD and FFT/PSD/ACF methods based on four classifiers with Recursive Feature Elimination (RFE). a) 5.08 cm (2 inch) overhang length b) 6.35 cm (2.5 inch) overhang length c) 8.89 cm (3.5 inch) overhang length d) 11.43 cm (4.5 inch) overhang length.}
\label{fig:compare_WPT_EEMD_SPM}
\end{figure}

\begin{figure}[H]
\centering
\includegraphics[width=0.9\textwidth,height=.95\textheight,keepaspectratio]{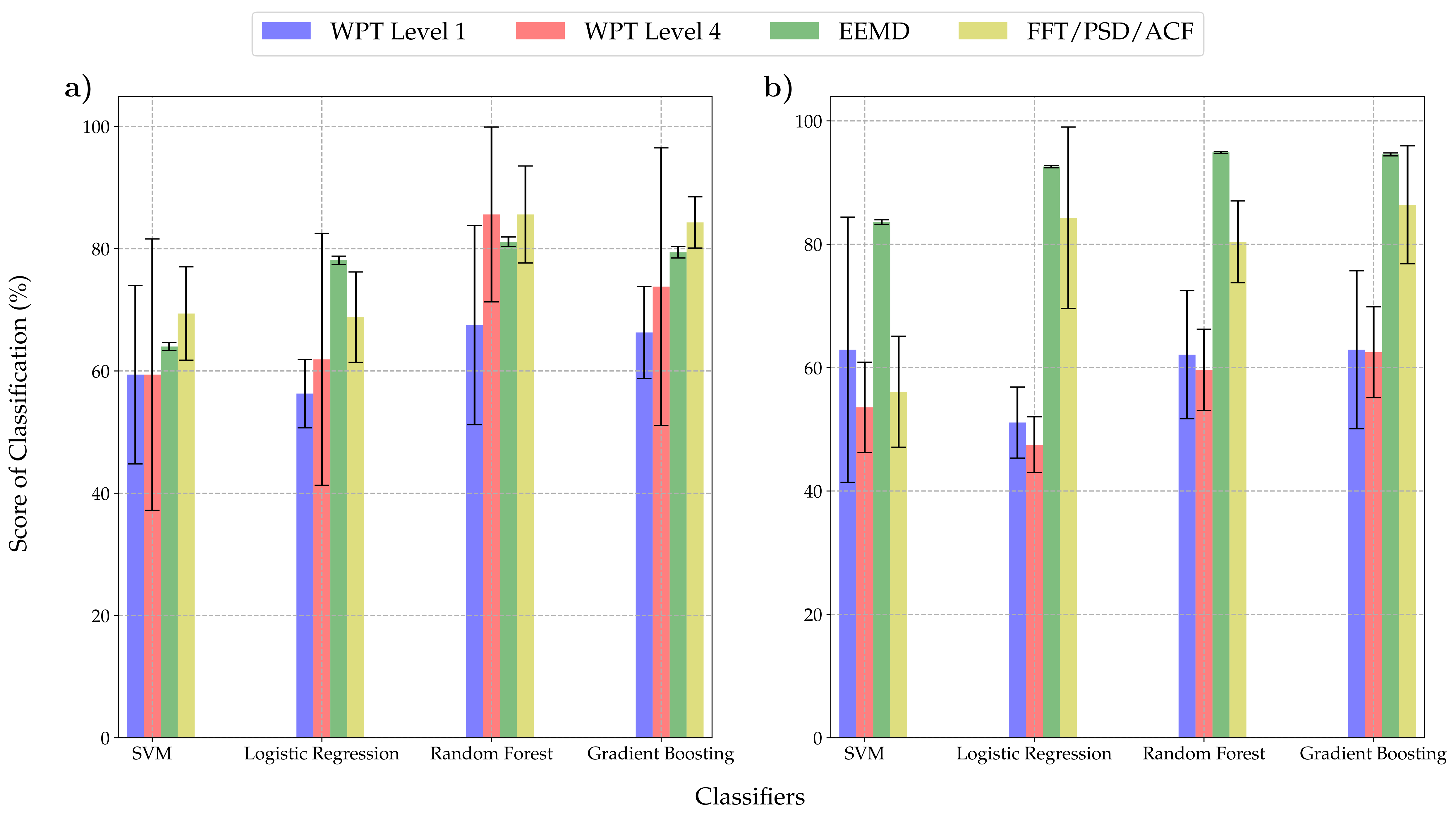}
\caption{Comparison of transfer learning results of Level 1 WPT, Level 4 WPT, EEMD and FFT/PSD/ACF methods based on four classifiers with Recursive Feature Elimination (RFE). 
a) Training set: 5.08 cm (2 inch) overhang length, Test set:11.43 cm (4.5 inch) overhang length
b) Training set: 11.43 cm (4.5 inch) overhang length, Test set:5.08 cm (2 inch) overhang length.}
\label{fig:comparison_high_acc_4_classifier_transfer_learning}
\end{figure}

Four different classifiers were utilized to compare traditional signal processing based feature extraction methods to the WPT/EEMD results obtained from Ref.~\cite{Yesilli2019}. 
The results obtained using these classifiers for each cutting configuration are provided in Table~\ref{tab:FFT_PSD_ACF}.
The table shows that classifiers trained with the traditional signal processing features have an overfitting problem (significantly higher accuracy rates when training versus testing) when RFE is not utilized.
Recursive feature elimination was used to solve this problem, see Section~\ref{sec:Feature_Extraction}. 
Table~\ref{tab:FFT_PSD_ACF} only reports the best results obtained from the classifiers for each overhang length. 
These results are also compared to the ones obtained with WPT and EEMD in Fig.~\ref{fig:compare_WPT_EEMD_SPM}.
It is seen that the FFT/PSD/ACF method has higher accuracies in Fig~\ref{fig:compare_WPT_EEMD_SPM}. 
On the other hand, that same method has significantly lower accuracies with respect to WPT/EEMD in transfer learning a classifier is trained using the 11.43 cm overhang length and is tested on the 5.08 cm overhang length (see Fig.~\ref{fig:comparison_high_acc_4_classifier_transfer_learning}b). 
For the first case in transfer learning, where the training set is 5.08 cm (2 inch) and the test set is 11.43 cm (2.5 inch), the FFT/PSD/ACF method shares the highest test accuracy with the Level 4 WPT albeit with smaller standard deviation.

%------------------------------
%------------------------------

\begin{table}[h]
\centering
\caption{Results obtained with signal processing feature extraction method for four overhang length with four different classifiers.}
\label{tab:FFT_PSD_ACF}
\resizebox{\textwidth}{!}{
\begin{tabular}{|l|c|c|c|c|c|c|c|c|}
\hline
\multicolumn{1}{|c|}{} & \multicolumn{4}{c|}{Without RFE} &  \multicolumn{4}{c|}{With RFE} \\
\hline
\multicolumn{1}{|c|}{} & \multicolumn{2}{c|}{5.08 cm (2 inch)} & \multicolumn{2}{c|}{6.35 cm (2.5 inch)} & \multicolumn{2}{c|}{5.08 cm (2 inch)} & \multicolumn{2}{c|}{6.35 cm (2.5 inch)}\\
\hline
\multicolumn{1}{|c|}{Classifier} & Test Set & Training Set & Test Set& Training Set & Test Set & Training Set & Test Set& Training Set\\
\hline
SVM &$\SI{46.2 \pm 6.0} {\percent}$&$\SI{100.0 \pm 0.0}{\percent}$&$\SI{38.3 \pm 13.0}{\percent}$&$\SI{100.0 \pm 0.0}{\percent}$&$\SI{61.5 \pm 9.7}{\percent}$&$\SI{69.6 \pm 8.3}{\percent}$&$\SI{71.7 \pm 10.7}{\percent}$&$\SI{86.0 \pm 6.6}{\percent}$\\
Logistic Regression&$\SI{78.5 \pm 12.8} {\percent}$&$\SI{100.0 \pm 0.0}{\percent}$&$\SI{76.7 \pm 8.2}{\percent}$&$\SI{100.0 \pm 0.0}{\percent}$&$\SI{83.8 \pm 8.0}{\percent}$&$\SI{82.3 \pm 5.8}{\percent}$&$\SI{81.7 \pm 13.8}{\percent}$&$\SI{89.0 \pm 8.3}{\percent}$\\
Random Forest&\cellcolor[rgb]{0.13,0.67,0.8}$\SI{93.8 \pm 3.1} {\percent}$&$\SI{100.0 \pm 0.0}{\percent}$&\cellcolor[rgb]{0.13,0.67,0.8}$\SI{86.7 \pm 10}{\percent}$&$\SI{100.0 \pm 0.0}{\percent}$&$\SI{94.6 \pm 4.9}{\percent}$&$\SI{98.5 \pm 1.9}{\percent}$&\cellcolor[rgb]{0.13,0.67,0.8}$\SI{95.0 \pm 7.6}{\percent}$&$\SI{100.0 \pm 0.0}{\percent}$\\
Gradient Boosting					&$\SI{84.6 \pm 6.9}{\percent}$&$\SI{100.0 \pm 0.0}{\percent}$&$\SI{76.7 \pm 17.0} {\percent}$&$\SI{100.0 \pm 0.0}{\percent}$&\cellcolor[rgb]{0.13,0.67,0.8}$\SI{96.2 \pm 3.9}{\percent}$&$\SI{100.0\pm 0.0}{\percent}$&$\SI{93.3 \pm 8.2} {\percent}$&$\SI{100.0 \pm 0.0}{\percent}$\\
\hline
\multicolumn{1}{|c|}{} & \multicolumn{4}{c|}{Without RFE} &  \multicolumn{4}{c|}{With RFE} \\
\hline
\multicolumn{1}{|c|}{} & \multicolumn{2}{c|}{8.89 cm (3.5 inch)} & \multicolumn{2}{c|}{11.43 cm (4.5 inch)}& \multicolumn{2}{c|}{8.89 cm (3.5 inch)} & \multicolumn{2}{c|}{11.43 cm (4.5 inch)}\\
\hline
SVM												&$\SI{62.0 \pm 24.4}{\percent}$&$\SI{100.0 \pm 0.0}{\percent}$&$\SI{43.8 \pm 17.0}{\percent}$&$\SI{100.0 \pm 0.0}{\percent}$&$\SI{82.0 \pm 10.8}{\percent}$&$\SI{95.6 \pm 5.4}{\percent}$ &$\SI{61.3 \pm 13.1}{\percent}$&$\SI{68.6 \pm 9.1}{\percent}$\\
Logistic Regression				&$\SI{76.0 \pm 12.0}{\percent}$&$\SI{100.0 \pm 0.0}{\percent}$&$\SI{76.3 \pm 10.4}{\percent}$&$\SI{100.0 \pm 0.0}{\percent}$&$\SI{78.0 \pm 14.0}{\percent}$&$\SI{88.9 \pm 8.6}{\percent}$ &$\SI{70.0 \pm 16.0}{\percent}$&$\SI{82.9 \pm 6.5}{\percent}$\\
Random Forest							&\cellcolor[rgb]{0.13,0.67,0.8}$\SI{90.0 \pm 9.4}{\percent}$&$\SI{100.0 \pm 0.0}{\percent}$&\cellcolor[rgb]{0.13,0.67,0.8}$\SI{90.0 \pm 9.4}{\percent}$&$\SI{100.0 \pm 0.0}{\percent}$&\cellcolor[rgb]{0.13,0.67,0.8}$\SI{92.0 \pm 9.8}{\percent}$&$\SI{100.0 \pm 0.0}{\percent}$ &\cellcolor[rgb]{0.13,0.67,0.8}$\SI{91.3 \pm 9.8 }{\percent}$&$\SI{99.3 \pm 2.1}{\percent}$\\
Gradient Boosting					&$\SI{86 \pm 23.7}{\percent}$&$\SI{100.0 \pm 0.0}{\percent}$&$\SI{83.8 \pm 11.3}{\percent}$&$\SI{100.0 \pm 0.0}{\percent}$&$\SI{86.0 \pm 9.2}{\percent}$&$\SI{100.0 \pm 0.0}{\percent}$ &$\SI{88.8 \pm 11.8}{\percent}$&$\SI{100.0 \pm 0.0}{\percent}$\\
\hline
\end{tabular}}
\end{table}

%------------------------------
%------------------------------

\begin{figure}[H]
\centering
\includegraphics[width=0.75\textwidth,height=.80\textheight,keepaspectratio]{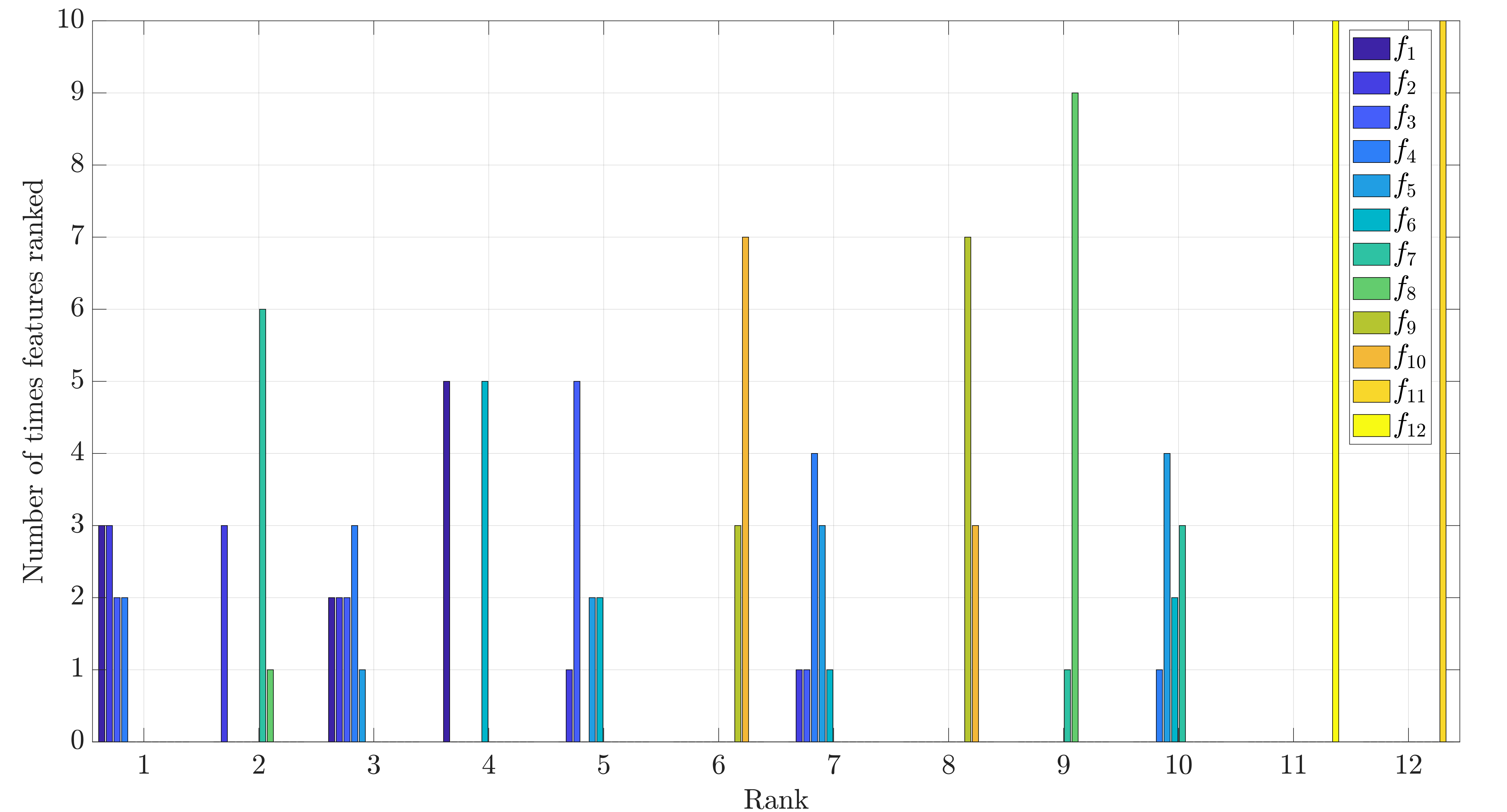}
\caption{Bar plot for feature ranking obtained from SVM-RFE for 5.08 cm (2 inch) overhang case with FFT/PSD/ACF based method. ($f_{1},\ldots,f_{4}$), ($f_{5},\ldots,f_{8}$) and ($f_{9},\ldots,f_{12}$) belong to features obtained from peaks of FFT, PSD and ACF respectively.}
\label{fig:Feature_Ranking_Bar_Plot_SVM_FFT}
\end{figure}

We explained the features extraction in Section~\ref{sec:Feature_Extraction}. 
Classification for this method has been performed in the same way we did for WPT/EEMD.
The left hand side of Table~\ref{tab:FFT_PSD_ACF} shows the classification results without using feature ranking. 
Note how the test accuracy is significantly lower than the training accuracy. 
This is a classical symptom of over fitting, i.e., using too many (unnecessary) features for training. 
In contrast, the right hand side of the same table shows that using feature ranking mitigated the over fitting problem. 
In addition, if we compare the results in Table~\ref{tab:FFT_PSD_ACF}, it is seen that the FFT/PSD/ACF based feature extraction methods have better accuracies than WPT and EEMD. 
Another approach to combat overfitting is to utilize Cross Validation (CV). Table \ref{tab:FFT_PSD_ACF_cv_results} provides the results obtained with CV where $10$-fold CV was used for the 5.08 cm (2 inch) and 11.43 cm (2.5 inch) cases, while 5-fold CV was used for the remaining cutting configurations. 
The results show that while CV somewhat mitigates the overfitting problem, it does not completely solve it. 
For example, there is overfitting for the classification accuracy of SVM for the results obtained without using RFE in Table \ref{tab:FFT_PSD_ACF}. 
Although CV decreased the difference between mean accuracies of test set and training set, there is still at least 20\% accuracy difference between training and test sets for overhand lengths 5.08, 6.35, and 11.43 cm (2, 2.5 and 4.5 inch). 
In addition, we can see a decrease in the standard deviation of the results in Table~\ref{tab:FFT_PSD_ACF_cv_results} in comparison to the ones in Table~\ref{tab:FFT_PSD_ACF}.
Figure~\ref{fig:compare_WPT_EEMD_SPM} provides bar plots for the classification accuracy for each cutting configurations along with the associating error bars.  
Figure~\ref{fig:Feature_Ranking_Bar_Plot_SVM_FFT} shows how many times each feature was ranked for the traditional signal processing based methods. 
This gives an indication of where the most highly ranked features come from. 
The figure shows that that most highly ranked features correspond to the FFT peaks, followed by PSD then ACF features.

%---------------
\begin{table}[h]
\centering
\caption{Cross validated results obtained with signal processing feature extraction method for four overhang length with four different classifiers.}
\label{tab:FFT_PSD_ACF_cv_results}
\resizebox{\textwidth}{!}{
\begin{tabular}{|l|c|c|c|c|c|c|c|c|}
\hline
\multicolumn{1}{|c|}{} & \multicolumn{4}{c|}{Without RFE} &  \multicolumn{4}{c|}{With RFE} \\
\hline
\multicolumn{1}{|c|}{} & \multicolumn{2}{c|}{5.08 cm (2 inch)} & \multicolumn{2}{c|}{6.35 cm (2.5 inch)} & \multicolumn{2}{c|}{5.08 cm (2 inch)} & \multicolumn{2}{c|}{6.35 cm (2.5 inch)}\\
\hline
\multicolumn{1}{|c|}{Classifier} & Test Set & Training Set & Test Set& Training Set & Test Set & Training Set & Test Set& Training Set\\
\hline
SVM 											&$\SI{74.2 \pm 4.7} {\percent}$&$\SI{94.8 \pm 4.7}{\percent}$&$\SI{58.3 \pm 3.1}{\percent}$&$\SI{98.5 \pm 3.1}{\percent}$&$\SI{90.0 \pm 1.9}{\percent}$&$\SI{89.8 \pm 1.9}{\percent}$&$\SI{73.3 \pm 6.5}{\percent}$&$\SI{75.2 \pm 6.5}{\percent}$\\
Logistic Regression				&$\SI{74.2 \pm 0.0} {\percent}$&$\SI{100.0 \pm 0.0}{\percent}$&$\SI{71.7 \pm 0.0}{\percent}$&$\SI{100.0 \pm 0.0}{\percent}$&$\SI{79.2 \pm 1.3}{\percent}$&$\SI{89.5 \pm 1.3}{\percent}$&$\SI{73.3 \pm 6.7}{\percent}$&$\SI{82.5 \pm 6.7}{\percent}$\\
Random Forest							&$\SI{89.2 \pm 0.0} {\percent}$&$\SI{100.0 \pm 0.0}{\percent}$&\cellcolor[rgb]{0.13,0.67,0.8}$\SI{93.3 \pm 0.0}{\percent}$&$\SI{100.0 \pm 0.0}{\percent}$&$\SI{91.7 \pm 1.3}{\percent}$&$\SI{95.2 \pm 1.3}{\percent}$&$\SI{93.3 \pm 0.0}{\percent}$&$\SI{100.0 \pm 0.0}{\percent}$\\
Gradient Boosting					&\cellcolor[rgb]{0.13,0.67,0.8}$\SI{91.7 \pm 0.0}{\percent}$&$\SI{100.0 \pm 0.0}{\percent}$&\cellcolor[rgb]{0.13,0.67,0.8}$\SI{93.3 \pm 0.0} {\percent}$&$\SI{100.0 \pm 0.0}{\percent}$&\cellcolor[rgb]{0.13,0.67,0.8}$\SI{97.5 \pm 0.0}{\percent}$&$\SI{100.0\pm 0.0}{\percent}$&\cellcolor[rgb]{0.13,0.67,0.8}$\SI{100.0 \pm 0.0} {\percent}$&$\SI{100.0 \pm 0.0}{\percent}$\\
\hline
\multicolumn{1}{|c|}{} & \multicolumn{4}{c|}{Without RFE} &  \multicolumn{4}{c|}{With RFE} \\
\hline
\multicolumn{1}{|c|}{} & \multicolumn{2}{c|}{8.89 cm (3.5 inch)} & \multicolumn{2}{c|}{11.43 cm (4.5 inch)}& \multicolumn{2}{c|}{8.89 cm (3.5 inch)} & \multicolumn{2}{c|}{11.43 cm (4.5 inch)}\\
\hline
SVM												&$\SI{86.7 \pm 0.0}{\percent}$&$\SI{100.0 \pm 0.0}{\percent}$&$\SI{41.7 \pm 7.0}{\percent}$&$\SI{88.4 \pm 7.0}{\percent}$&$\SI{80.0 \pm 5.8}{\percent}$&$\SI{91.1 \pm 5.8}{\percent}$ &$\SI{83.3 \pm 2.3}{\percent}$&$\SI{81.8 \pm 2.3}{\percent}$\\
Logistic Regression				&\cellcolor[rgb]{0.13,0.67,0.8}$\SI{93.3 \pm 0.0}{\percent}$&$\SI{100.0 \pm 0.0}{\percent}$&$\SI{46.7 \pm 6.6}{\percent}$&$\SI{94.0 \pm 6.6}{\percent}$&$\SI{86.7 \pm 3.6}{\percent}$&$\SI{92.9 \pm 3.6}{\percent}$ &$\SI{78.3 \pm 3.2}{\percent}$&$\SI{78.3 \pm 3.2}{\percent}$\\
Random Forest							&\cellcolor[rgb]{0.13,0.67,0.8}$\SI{93.3 \pm 0.0}{\percent}$&$\SI{100.0 \pm 0.0}{\percent}$&\cellcolor[rgb]{0.13,0.67,0.8}$\SI{91.7 \pm 0.0}{\percent}$&$\SI{100.0 \pm 0.0}{\percent}$&\cellcolor[rgb]{0.13,0.67,0.8}$\SI{93.3 \pm 0.0}{\percent}$&$\SI{100.0 \pm 0.0}{\percent}$ &\cellcolor[rgb]{0.13,0.67,0.8}$\SI{96.7 \pm 1.5}{\percent}$&$\SI{95.5 \pm 1.5}{\percent}$\\
Gradient Boosting					&$\SI{73.3 \pm 0.0}{\percent}$&$\SI{100.0 \pm 0.0}{\percent}$&$\SI{85.0 \pm 0.0}{\percent}$&$\SI{100.0 \pm 0.0}{\percent}$&$\SI{80.0 \pm 0.0}{\percent}$&$\SI{100.0 \pm 0.0}{\percent}$ &$\SI{91.7 \pm 0.0}{\percent}$&$\SI{100.0 \pm 0.0}{\percent}$\\
\hline
\end{tabular}}
\end{table}

%------------------------------
%---------------
\begin{table}[h]
\centering
\caption{Comparison highest accuracies obtained with four different classifier by applying of transfer learning results for 5.08 cm (2 inch) and 11.43 cm (4.5 inch) cases.}
\label{tab:transfer_learning_WPT_EEMD_FFT_best_4classifier}
\resizebox{0.8\textwidth}{!} {  % make the table a bit smaller so it doesn't spill out
\begin{tabular}{|l|c|c|c|c|c|c|c|c|}
\hline
\multicolumn{1}{|c|}{} & \multicolumn{4}{c|}{\shortstack{Training Set: 5.08 cm (2 inch) \\ Test Set: 11.43 cm (4.5 inch)}} & \multicolumn{4}{c|}{\shortstack{Training Set: 11.43 cm (4.5 inch) \\ Test Set: 5.08 cm (2 inch)}} \\
\hline
\makecell{Method} & SVM                    & \makecell{Logistic\\Regression} & \makecell{Random\\Forest}       &\makecell{Gradient\\Boosting}   & SVM                           & \makecell{Logistic\\Regression} & \makecell{Random\\Forest}    &\makecell{Gradient\\Boosting} \\
\hline
WPT Level 1				&$\SI{59.4}{\percent}$   &$\SI{56.3}{\percent}$   &$\SI{67.5}{\percent}$   &$\SI{66.3}{\percent}$   &$\SI{62.9}{\percent}$   & $\SI{51.1}{\percent}$   & $\SI{62.1}{\percent}$   & $\SI{62.9}{\percent}$ \\
WPT Level 4				&$\SI{59.4}{\percent}$   &$\SI{61.9}{\percent}$   &\cellcolor[rgb]{0.13,0.67,0.8}$\SI{85.6}{\percent}$   &\cellcolor[rgb]{0.63,0.79,0.95}$\SI{73.8}{\percent}$   &$\SI{53.6}{\percent}$   & $\SI{47.5}{\percent}$   & $\SI{59.6}{\percent}$   & $\SI{62.5}{\percent}$ \\
EEMD       				&$\SI{64.0}{\percent}$   &\cellcolor[rgb]{0.63,0.79,0.95}$\SI{78.1}{\percent}$   &\cellcolor[rgb]{0.63,0.79,0.95}$\SI{81.1}{\percent}$   &\cellcolor[rgb]{0.63,0.79,0.95}$\SI{79.4}{\percent}$   &$\SI{83.6}{\percent}$   & $\SI{92.6}{\percent}$   & \cellcolor[rgb]{0.13,0.67,0.8}$\SI{94.9}{\percent}$   & $\SI{94.6}{\percent}$ \\
FFT/PSD/ACF       &$\SI{69.4}{\percent}$   &$\SI{68.8}{\percent}$   &\cellcolor[rgb]{0.13,0.67,0.8}$\SI{85.6}{\percent}$   &\cellcolor[rgb]{0.63,0.79,0.95}$\SI{84.4}{\percent}$   &$\SI{56.1}{\percent}$   & $\SI{84.3}{\percent}$   & $\SI{80.4}{\percent}$   & $\SI{86.4}{\percent}$ \\
\hline
\end{tabular}
}
\end{table}

In Table~\ref{tab:transfer_learning_WPT_EEMD_FFT_best_4classifier} and Fig.~\ref{fig:comparison_high_acc_4_classifier_transfer_learning}, we present comparison of all the transfer learning results. 
In general, either WPT or EEMD has the highest accuracy compared to FFT/PSD/ACF in all applications of transfer learning. 
However, the first application of transfer learning (see Fig.~\ref{fig:comparison_high_acc_4_classifier_transfer_learning}a) for Level 4 WPT and FFT/PSD/ACF has equal test set scores obtained with Random Forest Classification.
For the transfer learning application where we train the classifier on 11.43 cm (4.5 inch) and test it on the 5.08 cm (2.5 inch), some of the classifiers provide higher accuracy for FFT/PSD/ACF than they do for WPT. 
However, the standard deviation of the FFT/PSD/ACF results is higher than those for WPT, and EEMD is the method with highest accuracy for all classifiers.

\section{Discussion}
\label{sec:Discussion}
Although FFT/PSD/ACF based methods have better classification results for our data set, these features have some drawbacks. 
The first issue is the over fitting problem when no feature ranking is used. 
The second challenge is the peak picking problem itself which is essential to the standard implementation of an FFT/PSD/ACF approach, see Section~\ref{sec:Feature_Extraction}.  
Choosing `true' peaks is not an easy task, and it involves tuning parameters such as the minimum peak height, the distance between peaks, etc. 
In addition, practically, these limits will need to be set once and and used for all incoming data streams.  
There is a high chance that the algorithm can miss some of the real peaks with these limitations. 
The third limitation is the poor transfer learning capability of features extracted using FFT/PSD/ACF as we explain below.  

We applied two types of transfer learning: in the first application we train on the 5.08 cm (2 inch) case and test on the 11.43 cm (4.5 inch) case. 
In the second application the training and testing sets for these two overhang cases are reversed.  
In the first application, we saw that the FFT/PSD/ACF shares the highest score with Level 4 WPT while it has considerably lower scores in the second application where the EEMD-based method provides the best scores. 
In each cutting configuration, we have different chatter frequencies and the places of the peaks in the graphs can differ significantly when the overhang length varies.  
This can lead to lower transfer learning accuracies for FFT/PSD/ACF based method.

\section{Conclusion}
\label{sec:Conclusion}
In this study, we applied traditional signal processing tools (FFT, PSD, and ACF) for feature extraction from cutting signals obtained from a turning experiment. 
We compared the results to the ones obtained from two widely used chatter detection methods: WPT and EEMD. 
Our results show that the classification results for FFT/PSD/ACF with feature ranking can provide higher test set accuracies for some of the classifiers, namely Random Forest Classification and Gradient Boosting. 
If we compare the overall best score for each overhang length, it is seen that the FFT/PSD/ACF based methods provide the best results.  
The application of transfer learning using featured obtained from FFT/PSD/ACF showed that the classification accuracy drops when training and testing using two different overhand cases. 
This can be explained by having different chatter frequencies for each cutting configuration. 
This results in significantly different locations of the frequency peaks for different overhang lengths, thus leading to a drop in the classification accuracy. 

Despite their shortcomings, traditional signal processing methods can yield highly-tuned classifiers with superior accuracy. 
This makes these methods suitable for manufacturing processes whose parameters are not expected to drift too much in comparison to the training data. 
In contrast, if the parameters of the underlying process are expected to significantly shift, then features based on FFT/PSD/ACF should be used with caution.

\section*{Acknowledgement}
This material is based upon work supported by the National Science Foundation under Grant Nos.~CMMI-1759823 and DMS-1759824 with PI FAK.

%
% \clearpage
\bibliographystyle{ieeetr}
\bibliography{MSEC2020}
\end{document}